\documentclass[a4paper,11pt]{article}
\pdfoutput=1 

\usepackage{jinstpub} 

\title{\boldmath 
Beam test characterisation of a Plastic Scintillator Prototype 
for the space-based cosmic ray experiment HERD
}

\author[a,1]{P.W. Cattaneo,\note{Corresponding author.}}
\author[b]{M. Pullia,}
\author[a]{M.C. Prata,}
\author[a]{A. Rappoldi}
\author[a]{and M. Rossella}

\affiliation[a]{INFN Pavia, Via Bassi 6, Pavia, Italy}
\affiliation[b]{CNAO, Str. Campeggi, 53, Pavia, Italy}

\emailAdd{paolo.cattaneo@pv.infn.it}

\abstract{
The High Energy cosmic-Radiation Detector (HERD) facility is planned to go
onboard China's Space Station, planned to be operational starting in around 
2025 for about 10 years. The main scientific objectives of HERD are the 
search for signals of dark matter annihilation products, precise cosmic
electron/positron spectrum and measurements of anisotropy up to 10 TeV, 
precise cosmic ray spectrum and composition measurements up to the knee 
energy (1 PeV), and high energy $\gamma$-ray monitoring and survey. 
HERD consists of a 3D cubic crystals calorimeter (CALO) surrounded by 
microstrip silicon trackers (STKs) and scintillating fiber trackers (FIT)
and by a Plastic Scintillator Detector (PSD) for $\gamma$-ray veto and 
ion charge measurement.
A PSD prototype consisting of a scintillator tile readout by two arrays of SiPMs
on opposite sides has been tested with proton and C ion beam at the CNAO
(Centro Nazionale Adroterapia Oncologica) in Pavia, (Italy).
Preliminary results on charge resolution are presented.
}

\keywords{Scintillators, Photon detectors for UV, visible and IR photons (SiPM),
          dE/dx detectors}

\proceeding{INSTR20: Instrumentation for Colliding Beam Physics \\
  24--28 February 2020,\\ Novosibirsk, Russia}

\begin{document}
\maketitle
\flushbottom

\section{The HERD experiment}
\label{sec:intro}

\begin{figure}[htbp]
\centering 
\includegraphics[width=.9\textwidth]{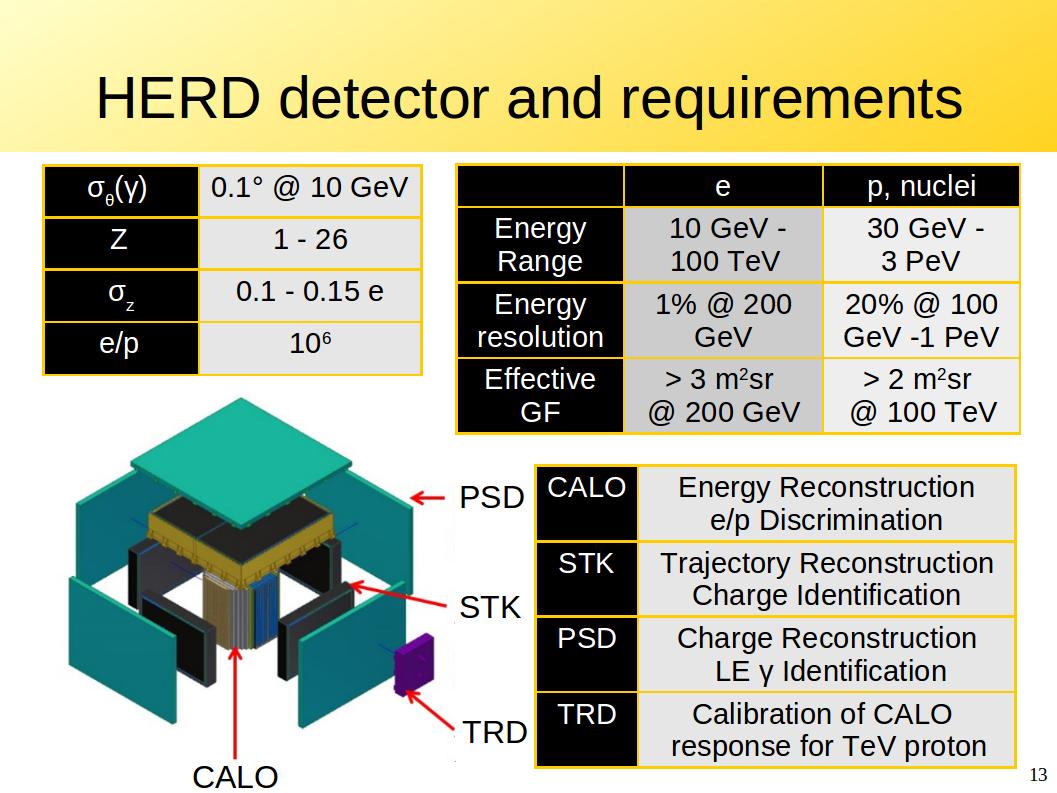}
\caption{\label{fig:herddetector} A sketch of the HERD detector.}
\end{figure}

\begin{figure}[htbp]
\centering 
\includegraphics[width=.9\textwidth,trim=30 110 0 0,clip]{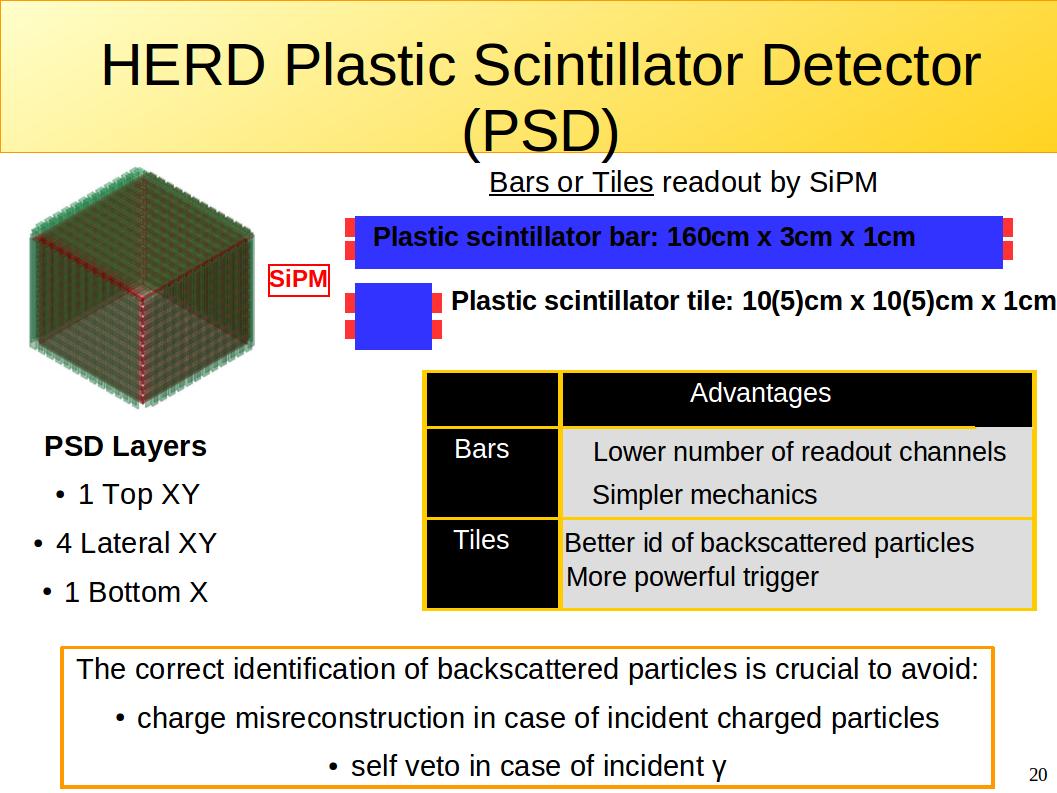}
\caption{\label{fig:psdherd} A sketch of the two possible solutions for the HERD PSD.}
\end{figure}

The High Energy cosmic-Radiation Detection (HERD) facility will be one of the space 
astronomy payloads on board the future Chinese space station. The goal of HERD 
is the direct detection of cosmic rays at the knee region (~ 1 PeV), with a detector 
able to measure electrons, photons and nuclei with high energy resolution 
(1\% for electrons and photons at 200 GeV and 20\%\ for nuclei at 100 GeV - PeV), 
a geometrical factor ten times larger than the one of present generation missions 
(> 2 m$^2$ sr), and life-time longer than ten years \cite{herd2018}. 
The primary objectives of HERD are the indirect search for dark matter particles and the 
precise measurement of energy distribution and composition of cosmic rays from 30 GeV 
up to a few PeV, determining the origin of the knee structure of the spectrum. 
Furthermore, HERD will monitor the high energy gamma-ray sky from $\sim 500$ MeV, 
observing gamma-ray bursts, active galactic nuclei, galactic microquasars, etc. 
HERD will be composed of a homogeneous calorimeter, surrounded by a particle tracker 
and a plastic scintillator detector (PSD) as shown in Fig.~\ref{fig:herddetector}.
The PSD consists of one or two layers of plastic scintillators surrounding the whole detector
segmented in bars or tiles readout by SiPMs. 
Its goal is to provide trigger and veto information and to measure the atomic
number Z of the ions through their energy loss.\\
The two mechanical solutions under investigations are sketched in Fig.~\ref{fig:psdherd} 
where the PSD can be segmented in tiles or bars. 
In the following we present the test of a prototype of the tile solution.

%

\section{The accelerator facility}

\begin{figure}[htbp]
\centering 
\includegraphics[width=.45\textwidth,clip]{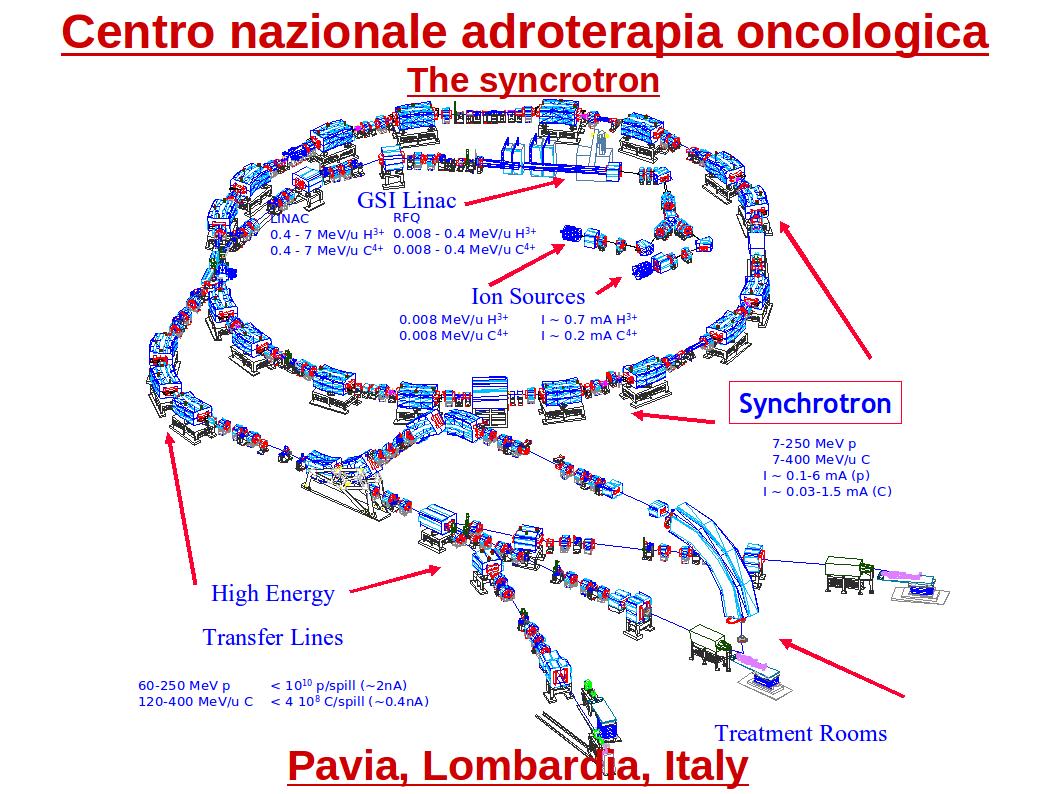}
\qquad
\includegraphics[width=.45\textwidth]{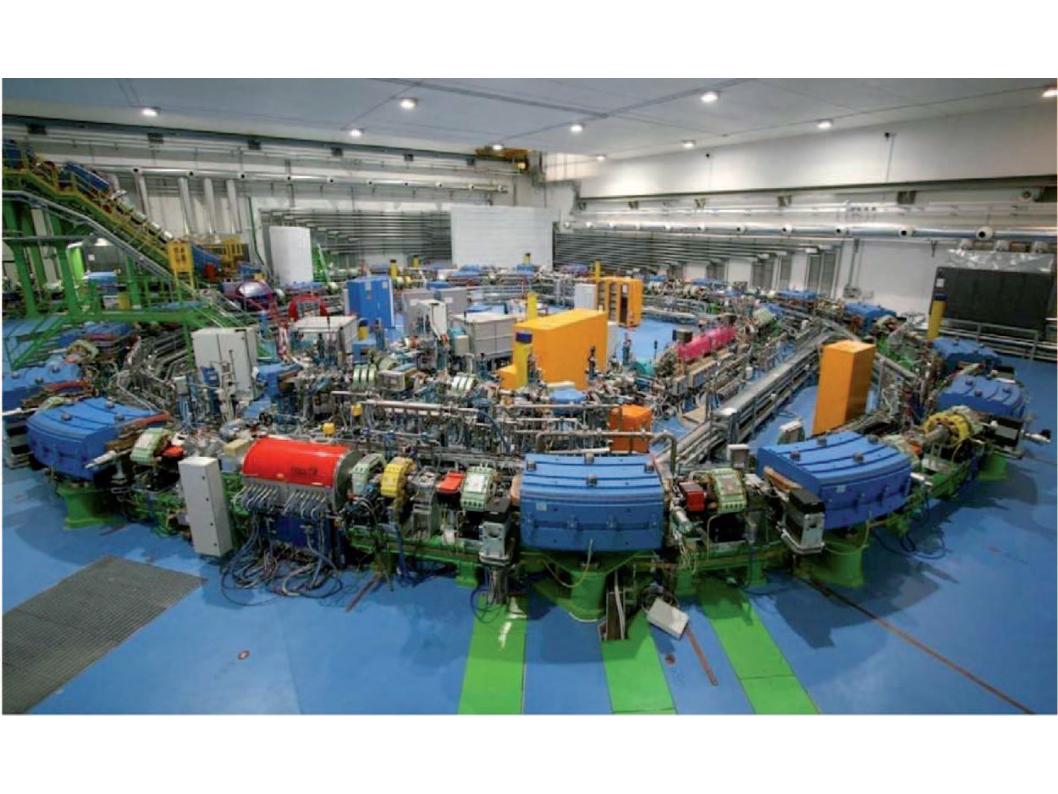}
\caption{\label{fig:cnao} A sketch of the CNAO synchrotron (left). A view of the CNAO synchrotron (right).}
\end{figure}

The CNAO \footnote{Centro Nazionale Adroterapia Oncologica, Pavia, Italy} \cite{cnao}
is a accelerator facility dedicated a hadron therapy to treat cancer, see Fig.\ref{fig:cnao}.\\
The beams available at CNAO are protons with kinetic energy in the continuous 
range 60-250 MeV and C ions with kinetic energy in the continuous range 120-400 MeV/u.
The rationale behind these ranges is that hadron therapy is particularly effective 
because of the large energy release at the Bragg peak for hadrons close to their 
stopping points.\\
The particle rate available is up to $10^{10}$ p/s or $4\times 10^8$ C/s and 
can be scaled down almost arbitrarily with appropriate support from the 
accelerator operators.
The beam spot is Gaussian in shape of size $\sigma_{x,y}=0.2-0.8$ cm depending 
on the energy and ion type.\\
In the following measurements the beam rate was tuned to a rate sufficiently low to 
have a negligible pileup. 

\begin{table}[htbp]
\centering
\caption{\label{tab:beamenergy} Ion types used in the test with beam kinetic energies per nucleon E$_\mathrm{k}$/A,
$\beta$ and $(Z/\beta)^2$ (see text).}
\smallskip
\begin{tabular}{|l|c|c|c|c|c|c|c|c|c|}
\hline
Beam type &\multicolumn{4}{c}{p} \vline & \multicolumn{5}{c}{C} \vline \\
\hline
E$_\mathrm{k}$/A (MeV) & 70    & 120   & 170   & 226   & 115   & 190   & 260   & 330   & 400 \\
$\beta$                & 0.366 & 0.462 & 0.532 & 0.592 & 0.454 & 0.555 & 0.622 & 0.672 & 0.713 \\
$(Z/\beta)^2$          & 7.46  & 4.68  & 3.53  & 2.85  & 174.6 & 116.9 & 93.0  & 79.7  & 70.8 \\
\hline
\end{tabular}
\end{table}

We exploited the full energy range for both ions; the energy values selected are shown in
Table~\ref{tab:beamenergy}.

\begin{figure}[htbp]
\centering 
\includegraphics[width=.9\textwidth,trim=30 110 0 0,clip]{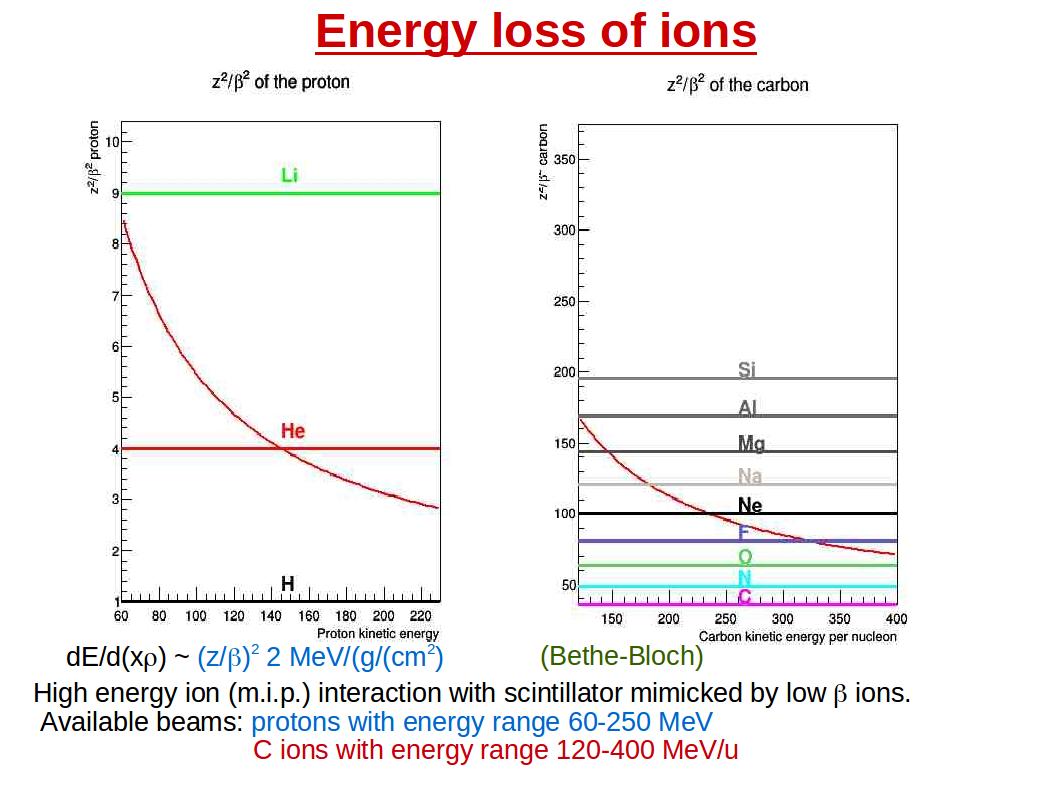}
\caption{\label{fig:ionloss} The $(Z/\beta)^2$ of protons (left) and carbon ions (right) versus kinetic
energy per nucleon compared with that of various relativistic ions.}
\end{figure}

The energy loss in the PSD prototype of ions with the energies in Table~\ref{tab:beamenergy}
is given by the Bethe-Bloch formula that describes the energy loss in MeV per length times density
\cite{pdg2018}

\begin{equation}
\label{eq:bethebloch}
\frac{dE}{d(x\rho)} \sim (Z/\beta)^2 2.0\, \mathrm{MeV/(g/cm^2)}
\end{equation}

where Z is the atomic number of the impinging ion and $\beta$ its velocity.
The energy loss is proportional to $(Z/\beta)^2$ so that the average loss due to 
relativistic ions equals that due to lighter ions with $\beta$ significantly smaller than one.\\
As shown in Fig.~\ref{fig:ionloss} the $(Z/\beta)^2$ values for the ions available at CNAO
cover the value of relativistic ions up to Al.

\section{The device under test}

\begin{figure}[htbp]
\centering 
\includegraphics[width=.45\textwidth]{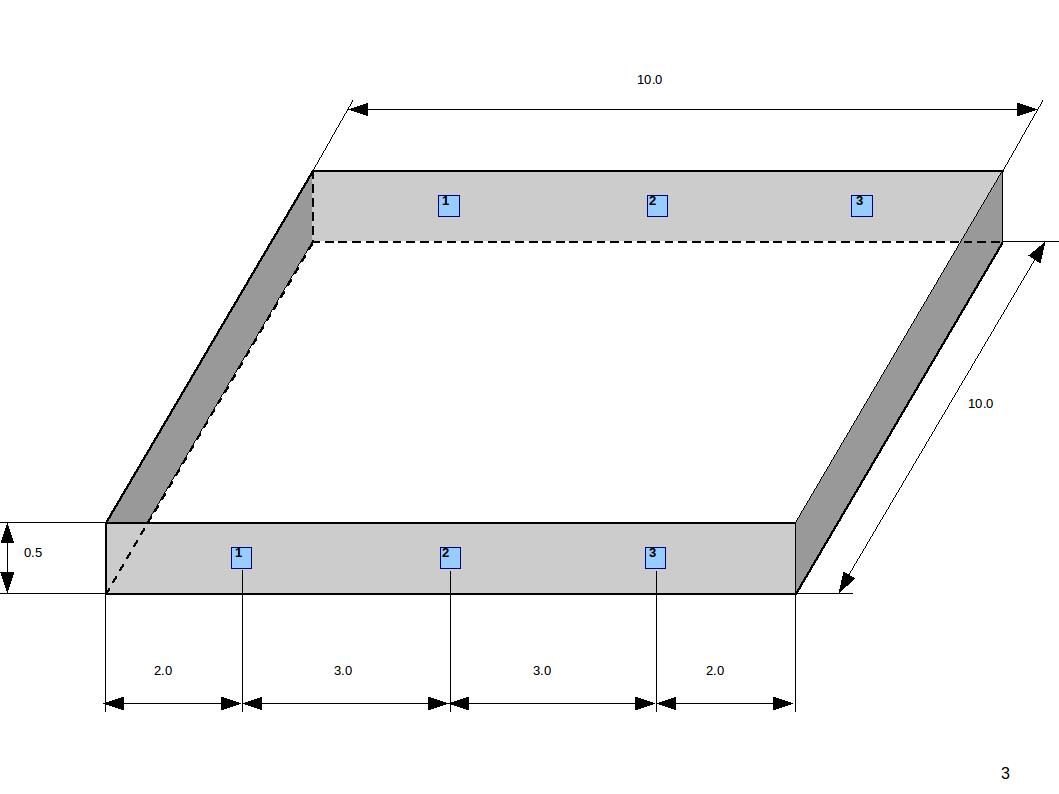}
\qquad
\includegraphics[width=.45\textwidth]{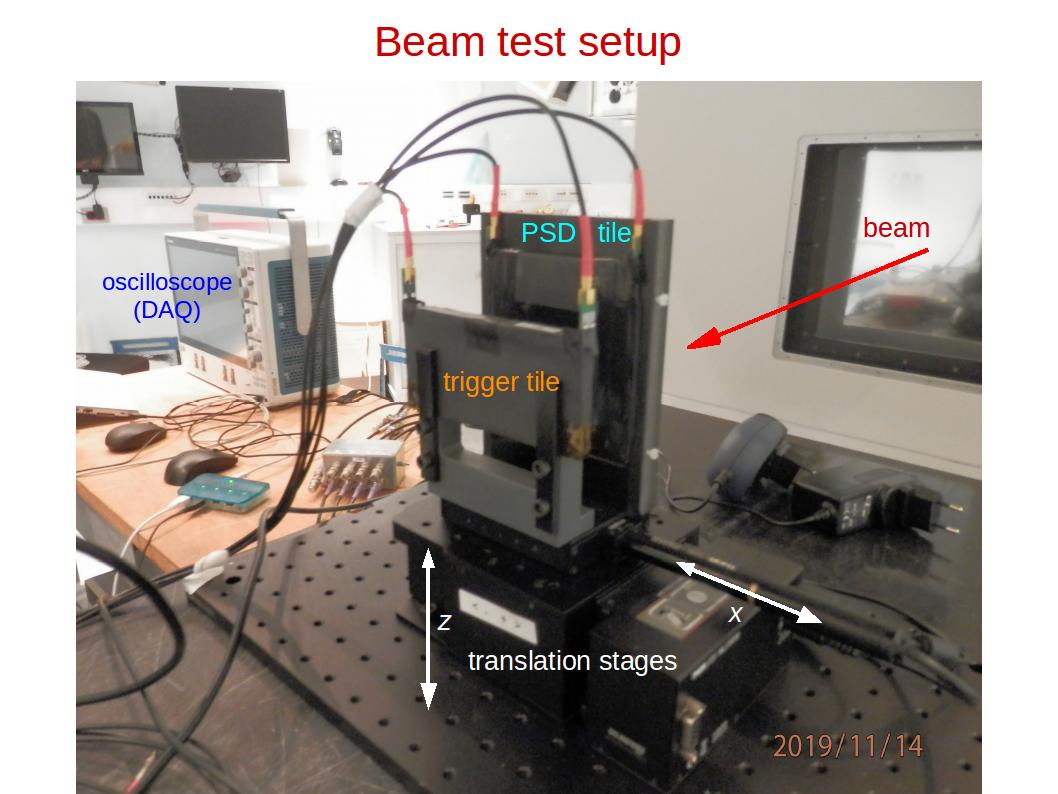}
\caption{\label{fig:tile} A sketch of the SiPMs positioning on the tile surface (left). 
        A picture of the device under test consisting of two tiles readout by SiPMs (right).}
\end{figure}

The goal of the beam test is to study the response of a tile of scintillator readout 
by SiPMs to hadrons impinging on it with the aim of measuring the discriminating power 
of different $Z/\beta$ and therefore of different relativistic ions.\\
In order to collect unbiased samples of events we built a setup consisting of two adjacent 
tiles shown in Fig.~\ref{fig:tile} (right). 
The scintillator tile marked as trigger tile is a spare tile of the MEG2 experiment timing 
counter \cite{meg2019} readout by two opposite sides by 6+6 SiPMs connected in parallel
that is used to trigger the data acquisition. 
The scintillator tile (EJ200) marked as PSD tile is the prototype for the Herd PSD: 
its size is $10\times 10$ cm$^2$ readout by 3+3 SiPM from Hamamatsu (S12572) positioned 
as shown in Fig.~\ref{fig:tile} (left). The SiPMs on each side are connected in parallel.
The two signals (right and left) from the PSD tile and the two signals (right and left) 
from the MEG2 tile are sent directly to a 
Tektronix MSO64 oscilloscope with four input channels sampling at 12.5 GHz with 12 bit ADCs.
The trigger is the sum of the amplitudes of the two signals from the MEG2 tile that is 
almost position independent.

\section{Results}

\begin{figure}[htbp]
\centering 
\includegraphics[width=.95\textwidth]{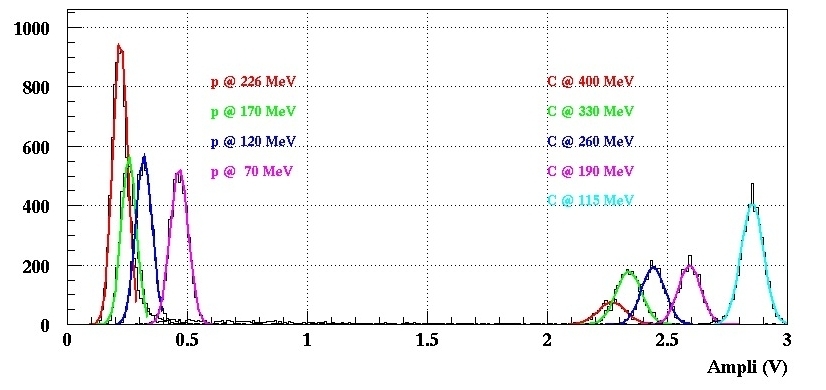}
\caption{\label{fig:hampl} Distribution of SiPM signal amplitudes for the different beam types and energies.}
\end{figure}

\begin{figure}[htbp]
\centering 
\includegraphics[width=.70\textwidth]{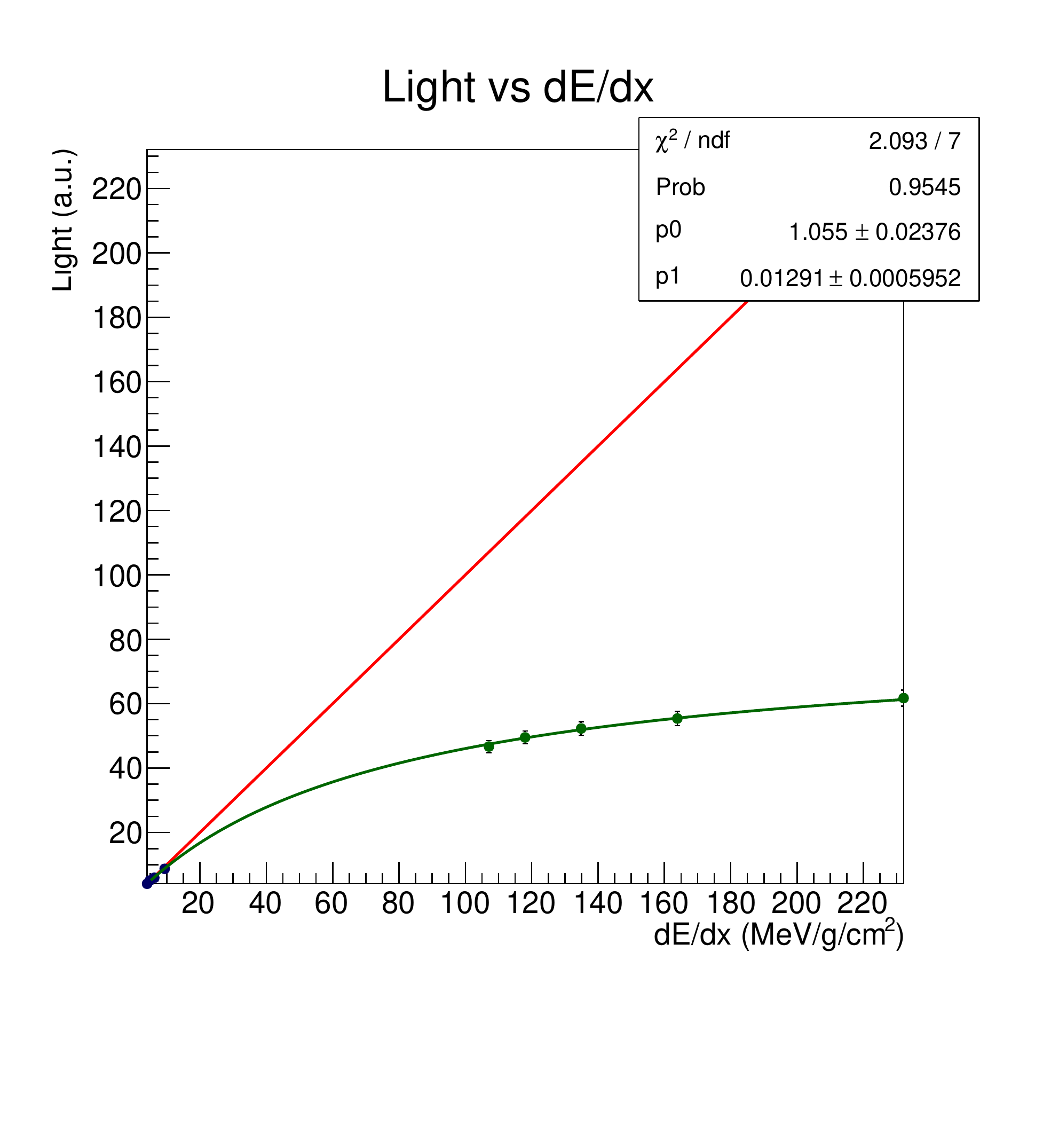}
\caption{\label{fig:birk} Light output versus the average energy loss. The red line 
  is the expected result in case of exact proportionality between energy loss and light output.
  Blue (p) and Green (C) axes report the corresponding kinetic energy per nucleon. 
  The green curve shows the result of the data fit with the Birks' formula (see text).}
\end{figure}

A first set of measurements is the light collected versus the expected energy loss in the tile.
The signals from the MEG2 tile are used only for triggering.
The relevant quantity is the sum, almost position independent, of the two signals from the HERD tile.
Figure~\ref{fig:hampl} reports the distribution of the sum amplitudes for all different 
beam types and energies.\\
In Fig.\ref{fig:birk} the amplitude versus the expected energy loss is displayed.
There is no linear relation as naively expected because of saturation of light production 
in the scintillator at high energy loss rate.
The data are fitted very well with the Birks' law \cite{birks}
\[
\frac{dL}{dx\rho} = S\frac{\frac{dE}{dx\rho}}{1 + k_\mathrm{B}\frac{dE}{dx\rho}}
\]
where $k_\mathrm{B}$ is a material dependent constant.
The result $k_\mathrm{B}=12.9\pm 0.6\times 10^{-3}$ g/cm$^2$/MeV is consistent with previous measurements
for plastic scintillators \cite{scinbk}.

\section{Conclusions}

We present some preliminary results on a beam test with p and C ions of a plastic scintillator (EJ200) tile.
The test performed at low $\beta$ covers a range of over 60 in expected average energy loss approximately
equivalent to the range from He to Al. The dynamic range in output amplitude is ~15 because of saturation effect
in production of scintillation light that is well described by Birks' law with a parameter $k_\mathrm{B}$ 
comparable to what measured in the past for similar materials.

\acknowledgments

We acknowledge the support and cooperation provided
by CNAO as the host laboratory.

\end{document}